\documentclass[12pt,preprint]{aastex}
\begin{document}

\title{A Resolved Circumstellar Disk around the Herbig Ae Star HD 100546
in the Thermal Infrared}
\author{Wilson M. Liu\footnote{Steward Observatory, University of Arizona,
933 N. Cherry Ave., Tucson, AZ, USA 85721, e-mail: wliu@as.arizona.edu},
Philip M. Hinz\footnotemark[1], Michael R. Meyer\footnotemark[1],
Eric E. Mamajek\footnotemark[1], and William F. Hoffmann\footnotemark[1], and
Joseph L. Hora\footnote{Harvard-Smithsonian Center for Astrophysics,
60 Garden St., MS 42, Cambridge, MA, USA 02138}}

\begin{abstract}
We present mid-infrared nulling interferometric and direct imaging observations
of the Herbig Ae star HD 100546 obtained with the Magellan I (Baade)
6.5 m telescope.  The observations show resolved circumstellar
emission at 10.3, 11.7, 12.5, 18.0, and 24.5 $\mu$m.  Through the nulling
observations (10.3, 11.7 and 12.5 $\mu$m), we detect a
circumstellar disk, with an inclination of $45 \pm 15$ degrees with respect
to a face-on disk, a semimajor axis position angle of $150 \pm 10$
degrees (E of N), and a spatial extent of about 25 AU.  The direct images
(18.0 and 24.5 $\mu$m) show evidence for cooler dust with a spatial extent of
30-40 AU from the star.  The direct images also show evidence for an
inclined disk with a similar position angle as the disk detected by
nulling.  This morphology is consistent with models in which
a flared circumstellar disk dominates the emission.  However, the similarity
in relative disk size we derive for different wavelengths suggests that the
disk may have a large inner gap, possibly cleared out by the formation of a
giant protoplanet.  The existence of a protoplanet in the system also
provides a natural explanation for the observed difference between
HD 100546 and other Herbig Ae stars.
\end{abstract}

\section{Introduction}
Circumstellar disks provide insight into the formation of planetary systems.
These disks are observed most readily around luminous pre-main-sequence (PMS)
stars.  Herbig Ae (HAE) stars, the evolutionary precursors to intermediate mass
main-sequence stars such as Vega, have been identified to have
infrared (IR) excess emission.  The source of the emission has been
hypothesized by \citet{hill92} and \citet{la92}
to originate from a geometrically thin, optically thick circumstellar disk,
with an optically thin inner region and a high accretion rate in order to
explain the observed spectral energy distribution (SED) of such stars.  An
alternative interpretation suggests that the emission may be a result of a
"dusty nebula" (or envelope), rather than a disk \citep{hart93}.  Recent
modelling has shown a likely possibility to be a
disk which flares vertically with increasing radius from the star
\citep{cg97,kh87}.  Other studies have found
that these disks may have a more complex structure, incorporating an inner
hole and heating of the inner wall of the disk to account for an excess in
near-IR emission \citep{ddn01}.  Another model incorporates an extended
spherical envelope surrounding a thin disk \citep{miro99}.  We refer the
reader to \citet{natta_ppiv} for an extensive review of recent results.
In general, observations of the circumstellar environments
of PMS stars are an important step in determining which models are most
representative of their true environment, as well as
understanding the evolution of protoplanetary disks into planetary systems.

The nearby ($\sim$100 pc) HAE star HD 100546 has been the focus of
several studies. \citet{malf98} characterized the spectrum
of the star in the IR,
and identified several spectral features indicative of silicate and polycyclic
aromatic hydrocarbon (PAH) species in the circumstellar environment.  They also
found features in the spectrum of HD 100546 to be very similar to those in
comet Hale-Bopp, indicating the presence of cometary material in the system,
and hypothesize that the system could harbor giant protoplanets to explain
the presence of crystalline silicates in the cometary material. A recent study
by \citet{bou03} found the spectrum of HD 100546 to be dramatically different
from other HAE stars, and propose a model with a circumstellar disk with
an inner gap of 10 AU and a giant protoplanet.
Three studies, \citet{grady,almm} and \citet{pwl},
used coronagraphic observations at near-IR wavelengths to image the dust
disk in scattered light and characterize its spatial structure.
These studies detect evidence of an inclined dust disk, and
are in good agreement as to its inclination
($\approx 40 \degr$ from face-on), and position angle of its semimajor axis
($130 \degr$ to $160 \degr$ E of N).  Extended emission has also
been detected at 3.4 mm \citep{wilner}, and far-ultraviolet observations
of warm molecular hydrogen are also consistent with the presence of an
inclined disk \citep{lec03}.  The presence of circumstellar emission from HD
100546, as well as its relative proximity, make it an ideal target for nulling
interferometry.

Nulling interferometry is a technique used to study circumstellar environments
by suppressing the starlight which normally overwhelms any signal from the
faint circumstellar material.  The technique is implemented by
overlapping the pupils of two telescopes (or two subapertures from a single
telescope) with an appropriate path length difference to destructively
interfere the light.  The result is a sinusoidal transmission pattern where
the unresolved central point source is suppressed and the surrounding resolved
structure can be detected.

In this Letter, we present the results of nulling interferometric and direct
imaging observations of the HAE star HD 100546 in the mid-IR.  We
discuss the detection and structure of resolved emission surrounding the star
at several wavelengths.  For further background we refer the reader to
\citet{Hinz01}.

\section{Observations \& Data Reduction}
Observations were made in August 2001 and May 2002 at the Magellan I (Baade)
6.5 m telescope at Las Campanas Observatory near La Serena, Chile.  The
BracewelL Infrared Nulling Cryostat (BLINC) was the nulling interferometer
used to provide suppression of starlight, creating an interferometer with two
2.5 m subapertures and a baseline of 4 m.  The Mid-InfraRed Array Camera
\citep[MIRAC]{hoff98} provided the final stop
for the two beams of the interferometer.  Nulling observations
were taken at 10.3, 11.7 and 12.5 $\mu$m with 10\% bandpass at each wavelength.
Images of the science object, HD 100546, were taken at seven different
rotations of the interferometer baseline with respect to the sky in order to
probe the geometry of the circumstellar dust (discussed in
Section \ref{sec-null}).  Thirty-seven sets of 500 frames were taken (2 sets at
each combination of wavelength and rotation for all but one of the combinations),
each frame with an integration time of 0.5 s.  These observations were
interlaced with observations of calibrator stars, with the same integration
times.  Frames were sky subtracted using off-source sky frames taken in between
observations.  A custom IDL program was used to perform aperture photometry on
each image and pick out, for each set of 500 frames, the image with the best
instrumental null.

Direct images at 11.7, 18.0 and 24.5
$\mu$m were obtained in August 2001 and March 2003.  Images were taken with
a 3 Hz chop of $8 \arcsec$ in the horizontal direction of the detector, and
a nod of $8 \arcsec$ in the vertical direction after every 15 s of integration.
The 11.7 $\mu$m images
were taken to verify extended emission detected by nulling, with a total
integration time of 60 s.  The 18.0 and
24.5 $\mu$m images were taken with the purpose of detecting resolved
material at longer wavelengths and characterizing emission from cooler dust.
Total integration times were 170 s at 18.0 $\mu$m and 210 s at 24.5
$\mu$m.  Aperture photometry was performed on the direct images of HD 100546
and calibrator stars and relative fluxes for the science object were
transformed to absolute fluxes using calibrator fluxes taken from \citet{gez93}.
The absolute fluxes are 67 Jy at 11.7 $\mu$m, 123 Jy at 18.0 $\mu$m, and
165 Jy at 24.5 $\mu$m, and are good to 10\%.

\section{Results}
\subsection{Nulling Observations} \label{sec-null}
Table \ref{tab:obs} shows the source nulls
and errors achieved at each combination of
wavelength and rotation.  The source null for the science object
is calculated by subtracting the null achieved on the calibrator star from the
instrumental null achieved on the science object, and represents the percentage
of light remaining in the image when nulled, compared to the full flux when
the pupils are constructively interfered.  The best source null is calculated
for each set of 500 frames.  The adopted values for the source null
presented in Table \ref{tab:obs} are taken to be the average of the best nulls
in each of the two sets of frames taken at each combination of wavelength and
rotation. The error in the null is taken to be the difference in the the best
nulls in the two sets of frames.

The non-zero source nulls on HD 100546 are indicative of extended
emission surrounding the star.  Our results show that the source null of the
object varies as a function of rotation of interferometer baseline with
respect to the sky (Figure \ref{fig:nullplot}).  At all three
wavelengths, the null varies by a factor of 2 or more and appears to
have roughly the same dependence.  From this dependence, we can infer the
presence of an inclined structure, as well as its size and orientation in the
following manner.  The transmitted signature of the nulling interferometer is
an interference pattern with interference fringes along the
baseline.  If these fringes are parallel to the major axis of a disk, more
of the disk's light will be nulled, resulting in a lower percentage of
remaining light.  When the fringes are aligned orthogonally to the major axis,
the value of the source null will be higher (i.e., there is more light
remaining).  Therefore, there should be a variation in null with respect to
rotation of the baseline.  This relation will have a sinusoidal
form, with a period of $180 \degr$ and an amplitude dependent on the projected
emitting surface of the disk, and consequently its inclination.  We have
performed a least squares fit of such a function for the expected null
(see Fig. \ref{fig:nullplot}), $N = a + b * sin(PA + \theta)$.
The value of $a$ is a vertical offset (physically related to the size
of the disk), $b$ is the amplitude of the sine function (related
to the inclination of the disk), PA is the position angle of the semimajor
axis, and $\theta$ is the rotation of the interferometer baseline
relative to the sky.  We find the parameters shown in Table \ref{tab:sinefit}
as the best fits for the data.  The values derived for all three wavelengths
are in agreement with one another, with average derived
uncertainties in the nulls of 6.3, 5.5, and 5.5\% for 10.3, 11.7, and 12.5
$\mu$m, respectively.

In order to interpret these fits physically,
we have calculated the physical parameters which would yield these fit
parameters for two simple disk distributions: an inclined Gaussian disk and
a ring. Table \ref{tab:physfit} shows the best fit physical parameters for the
nulling data. The best fit position angle of the semimajor axis of the disk
is $140 \degr$ to $160 \degr$ (E of N), which is in agreement with
the values derived from near-IR coronagraphic studies
\citep{grady,almm,pwl}.  The inclination of the disk at 10.3 and 11.7 $\mu$m
is derived to be $30 \degr$ to $40 \degr$ from face-on, also in
agreement with the aforementioned studies, but these fits are only
marginally better than a face-on disk when comparing the
reduced $\chi^{2}$ ( = 1.5 and 2.2 for the 10.3 and 11.7 $\mu$m fits,
respectively) of the fits.  The 12.5 $\mu$m data show a larger amplitude in
the variation of the null, hence the inclination derived is greater, $\approx
60 \degr$ from face-on.  In this case, there is significant inclination of the
disk, as the fit yields reduced $\chi^{2} = 2.4$ as opposed to 7.5 for a
face-on disk.

The multiwavelength nature of our observations
also allow us to probe differences between the
distribution of different species (silicates, PAH, etc.) and the thermal
continuum.  Our observations at 10.3 $\mu$m and 11.7 $\mu$m probe
emission from silicates and PAH species, while the contiuum emission is roughly
probed by the 12.5 $\mu$m band \citep{malf98}, although the bandpass of this
filter may result in significant emission from PAH and silicates.  Our results
indicate that the emitting structure is more inclined at 12.5 $\mu$m than at
the other two wavelengths.  This suggests that emission from the
thermal continuum may have a more inclined structure than the flux from
emission lines of silicates and PAH.

\subsection{Imaging}
The 11.7 $\mu$m images verify the presence of resolved emission detected
in the nulling data.  The images of HD 100546 show an average full-width at
half-maximum (FWHM) of $\sim 0.5\arcsec$, while the calibrator star shows a
FWHM about 20\% smaller.  This implies a disk size of about 30 AU, which
confirms the disk sizes derived from the nulling data.

The 18.0 and 24.5 $\mu m$ direct images show evidence for extended emission
as well, with the FWHM values for HD 100546 images on average about 8 - 10\%
larger than those of the calibrator stars.
In order to determine the spatial extent of the extended emission, we
constructed an artificial source by convolving an artificial face-on
disk signature in the form of a two-dimensional Gaussian, with the PSF
from the calibrator star.  The artificial image was subtracted from the
actual image of HD 100546.  The width of the artificial Gaussian disk
was varied in steps of 0.1 pix (equivalently $0.012 \arcsec$, or 1.2 AU at
100 pc).  We adopt the disk size which resulted in the smallest
residual when subtracted from the image of HD 100546.  Figure \ref{fig:sub}
shows a typical image before and after subtractions of the artificial source,
plotted with the same greyscale.
The top-left image shows HD 100546 at 24.5 $\mu$m in an unsubtracted image.
The top-right image shows the subtracted frame with the
smallest residual.  The center frames shows the subtraction residuals
where the Gaussian disk was about 0.5 pix (FWHM) too small (left),
and too large (right).  Table \ref{tab:cooldisk} shows the
results of the model fitting at each wavelength.  The disk size
adopted for each wavelength is the average of the sizes determined from the
observations at the two epochs, with the error bars
adopted as the difference in sizes derived for the two epochs.

At these wavelengths, we are probing both the thermal continuum and emission
from silicates.  As expected from cooler dust, the 24.5 $\mu$m
emission extends farther out than the 18.0 $\mu$m dust.  We do
note the 24.5 $\mu$m band contains a strong emission line from silicates
\citep{malf98} that may contribute significantly on top of the thermal
emission.  The images also show that there may be
evidence for an inclined disk in the subtracted images, as the best
subtracted image still shows a roughly symmetric oversubtraction above and
below the center of the star.  In images at both
wavelengths and both epochs, the residuals showed this type of symmetric
structure, with peak of the positive residuals on a line orthogonal to the
trough of the negative residuals (see center panels, Fig. \ref{fig:sub}).
This structure in the residuals would be expected if the image of HD 100546
was slightly elliptical, perhaps as a result of a resolved inclined disk.
We attempted subtractions with artificial sources incorporating an inclined
(and rotated) Gaussian disk, rather than a face-on disk.  The
bottom panel of Figure \ref{fig:sub} shows the outcome of this subtraction,
resulting in a smaller residual than the best
subtraction with a face-on disk (top-right panel).  The position angle of
the artificial disk was changed in steps of $10 \degr$, and the inclination
was varied in steps of $5 \degr$ in order to determine the
orientation of the artificial disk which resulted in the smallest residual when
subtracted from the image of HD 100546.  The position angle of the semimajor
axis was found to be between $130 \degr$ and $170 \degr$, and the inclination
was $30 \degr$ to $40 \degr$ from
face-on, all roughly consistent with the orientation derived by previous
studies in the near-IR \citep{almm,pwl,grady}.

\section{Discussion} \label{sec-disc}
We are confident that we have resolved circumstellar emission from HD
100546 at all wavelengths probed in our observations.
We wish to compare the physical parameters we have derived for
this emission to current models for the circumstellar environments of Herbig
Ae stars.  Recent models we consider include those described in the
Introduction.  While observations at all three
wavelengths show evidence for an inclined disk, our observations at
10.3 and 11.7 $\mu$m are also consistent with a face-on
or spherical emitting body.  However, the 12.5 $\mu$m null variation does
provide convincing evidence for an inclined disk.  Furthermore, the
derived sizes at these wavelengths are increasing with increasing wavelength
(equivalently, decreasing temperature), as one might expect.
Somewhat puzzling are the derived sizes of the 18.0 and 24.5 $\mu$m disks. One
would expect the thermal emission at these longer wavelengths to be spatially
several times larger than the emission at the shorter wavelengths.  If the
source of the emission is a continuous flared disk, this relation would be
given by T $\sim\ r^{-0.5}$ \citep{cg97}.  However, the disks at 18.0 and 24.5
$\mu$m are only marginally larger than at shorter wavelengths.  This
discrepancy suggests that a continuous disk
that extends all the way into the dust sublimation
radius may not be an accurate model for the dectected emission.  Instead we
prefer a model with a large inner disk gap, possibly cleared out by a giant
protoplanet, as suggested in \citet{bou03}.
This would result in the shorter wavelength emission being detected further
from the star than expected from a continuous disk, and make the relative sizes of the
10 and 20 $\mu$m disks more similar than expected from a T $\sim\ r^{-0.5}$
relation.

The dectection of a disk around HD 100546 is also interesting in the context
of the observations of \citet{Hinz01}.  The previous study performed nulling
interferometric observations of three other HAE stars, HD 150193,
HD 163296, and HD 179218 and found that none of them had resolved
emission.  This placed an upper limit on the size of the 10.3 $\mu$m disk
of 20 AU.  This suggests that HD 100546 differs from these other PMS objects,
as it appears to have a larger disk at 10.3 $\mu$m.  A disk such as the one
observed around HD 100546 would have been resolvable around the three
stars observed by \citet{Hinz01}, and leads us to
conclude that the physical structure and/or composition of the circumstellar
environment is different in HD 100546 than the HAE stars.  This result
is consistent with the finding of \citet{bou03} that
the SED of HD 100546 is dissimilar to that of other HAE stars.
We also note that the hypothesis of giant protoplanet in the HD 100546 system
would provide a natural explanation for the difference between HD 100546 and
the stars observed in \citet{Hinz01}.
A full analysis of a larger sample of HAE stars is needed to confirm this
conclusion, and will be presented in a future paper.

\section{Acknowledgments}
WL acknowledges support from the Michelson Graduate Fellowship.
MM and EM acknowledge support through NASA
contract 1224768 administered through JPL. EM also thanks the NASA Graduate
Student Researchers Program (NGT5-50400) for support.  The authors
thank the staff of Las Campanas Observatory and the Magellan Project for
wonderful support.  BLINC was developed under a NASA/JPL grant for TPF.  MIRAC
is supported by grant AST 96-18850 from the NSF with additional support from
SAO.

\clearpage
\begin{deluxetable}{cccc}
\tablecaption{Source Nulls for HD 100546 \label{tab:obs}}
\tablewidth{0pt}
\tablehead{
\colhead{Rotation ($\degr$)} &
\colhead{10.3 $\mu$m} &
\colhead{11.7 $\mu$m} &
\colhead{12.5 $\mu$m}
}
\startdata
-80 & $36.9 \pm 1.6$ & $33.6 \pm 1.6$ & $29.9 \pm 3.3$\\
-77 & $33.8 \pm 5.0$ & $32.8 \pm 3.3$ & no data\\
-50 & $31.3 \pm 7.0$ & $19.7 \pm 3.6$ & $43.2 \pm 0.9$\\
-24 & $33.3 \pm 2.8$ & $28.8 \pm 3.2$ & $24.8 \pm 1.7$\\
+10 & $21.2 \pm 8.9$ & $23.2 \pm 3.3$ & $20.3 \pm 1.6$\\
+13 & $27.9 \pm 1.5$ & $25.0 \pm 1.3$ & no data\\
+40 & $19.1 \pm 0.4$ & $14.5 \pm 3.0$ & $6.0 \pm 6.3$\\
\enddata
\end{deluxetable}

\clearpage
\begin{deluxetable}{cccc}
\tablecaption{Best fit Sine Function Parameters \label{tab:sinefit}}
\tablewidth{0pt}
\tablehead{
\colhead{Parameter} &
\colhead{10.3 $\mu$m} &
\colhead{11.7 $\mu$m} &
\colhead{12.5 $\mu$m}
}
\startdata
a & 0.287 & 0.253 & 0.231\\
b & 0.070 & 0.045 & 0.148\\
PA (E of N) & 147 & 154 & 141\\
\enddata
\end{deluxetable}

\clearpage
\begin{deluxetable}{ccccc}
\tablecaption{Physical Parameters of Disk from Nulling Data \label{tab:physfit}}
\tablewidth{0pt}
\tablehead{
\colhead{$\lambda$ ($\mu$m)} &
\colhead{FWHM (Gaussian)} &
\colhead{Inclination (Gaussian)} &
\colhead{Diameter (Ring)} &
\colhead{Incl. (Ring)}
}
\startdata
10.3 & 24 AU & $40 \degr$ & 26 AU & $37 \degr$\\
11.7 & 25 & 34 & 27 & 32\\
12.5 & 30 & 63 & 33 & 60\\
\enddata
\end{deluxetable}

\clearpage
\begin{deluxetable}{cccc}
\tablecaption{Results from Direct Imaging \label{tab:cooldisk}}
\tablewidth{0pt}
\tablehead{
\colhead{$\lambda$ ($\mu$m)} &
\colhead{Disk FWHM (pix)} &
\colhead{Angular size (arcsec)} &
\colhead{Physical size (AU)}
}
\startdata
18.0 & $2.85 \pm\ 0.20$ & $0.34 \pm 0.02$ & $34 \pm 2$\\
24.5 & $3.55 \pm\ 0.20$ & $0.43 \pm 0.02$ & $43 \pm 2$\\
\enddata
\end{deluxetable}

\clearpage

\figcaption{Source null vs. Rotation of the interferometer baseline for 10.3,
11.7, and 12.5 $\mu$m.
\label{fig:nullplot}}

\figcaption{\emph{Top-left}: 24.5 $\mu$m image of HD 100546, 60 s integration.
         \emph{Top-right}: Residual from the best
	 subtraction of an artificial image (with a face-on disk).
	 \emph{Center-left}: Residual resulting
	 from an artificial disk 0.5 pix (FWHM) too small. \emph{Center-right}:
	 Residual from an artificial disk 0.5 pix too large.
	 \emph{Bottom}: Residual from a subtraction of an artificial source
	 with an inclined disk.	
\label{fig:sub}}


\begin{thebibliography}

\bibitem[Augereau et al.(2001)]{almm} Augereau, J.~C., Lagrange, A.~M.,
Mouillet, D., \& M{\' e}nard, F.\ 2001, \aap, 365, 78

\bibitem[Bouwman et al.(2003)]{bou03}
Bouwman, J., de Koter, A., Dominik, C., \& Waters, L.~B.~F.~M.\ 2003, \aap,
401, 577

\bibitem[Chiang \& Goldreich(1997)]{cg97} Chiang, E.~I.~\&
Goldreich, P.\ 1997, \apj, 490, 368

\bibitem[Dullemond, Dominik, \& Natta(2001)]{ddn01}
Dullemond, C.~P., Dominik, C., \& Natta, A.\ 2001, \apj, 560, 957

\bibitem[Gezari et al.(1993)]{gez93}
Gezari, D.~Y., Schmitz, M., Pitts, P.~S., \& Mead, J.~M.\ 1993,
Catalog of Infrared Observations, 3rd ed., NASA Reference Publication 1294.

\bibitem[Grady et al.(2001)]{grady} Grady, C.~A.~et al.\
2001, \aj, 122, 3396

\bibitem[Hartmann, Kenyon, \& Calvet(1993)]{hart93} Hartmann,
L., Kenyon, S.~J., \& Calvet, N.\ 1993, \apj, 407, 219

\bibitem[Hillenbrand et al.(1992)]{hill92}
Hillenbrand, L.~A., Strom, S.~E., Vrba, F.~J., \& Keene, J.\ 1992, \apj,
397, 613

\bibitem[Hinz, Hoffmann, \& Hora(2001)]{Hinz01} Hinz, P.~M.,
Hoffmann, W.~F., \& Hora, J.~L.\ 2001, \apjl, 561, L131

\bibitem[Hoffmann et al.(1998)]{hoff98} Hoffmann, W.~F., Hora,
J.~L., Fazio, G.~G., Deutsch, L.~K., \& Dayal, A.\ 1998, \procspie, 3354,
647

\bibitem[Kenyon \& Hartmann(1987)]{kh87} Kenyon, S.~J.~\&
Hartmann, L.\ 1987, \apj, 323, 714

\bibitem[Lada \& Adams(1992)]{la92} Lada, C.~J.~\& Adams,
F.~C.\ 1992, \apj, 393, 278

\bibitem[Lecavelier des Etangs et al.(2003)]{lec03}
Lecavelier des Etangs, A.~et al.\ 2003, \aap, 407, 935

\bibitem[Malfait et al.(1998)]{malf98} Malfait, K., Waelkens,
C., Waters, L.~B.~F.~M., Vandenbussche, B., Huygen, E., \& de Graauw,
M.~S.\ 1998, \aap, 332, L25

\bibitem[Miroshnichenko et al.(1999)]{miro99} Miroshnichenko, A.,
Ivezi{\' c} , {\v Z}., Vinkovi{\' c} , D., \& Elitzur, M.\ 1999,
\apjl, 520, L115

\bibitem[Natta, Grinin, \& Mannings(2000)]{natta_ppiv} Natta, A.,
Grinin, V., \& Mannings, V.\ 2000, Protostars and Planets IV, 559

\bibitem[Pantin, Waelkens, \& Lagage(2000)]{pwl} Pantin,
E., Waelkens, C., \& Lagage, P.~O.\ 2000, \aap, 361, L9

\bibitem[Wilner et al.(2003)]{wilner} Wilner, D.~J., Bourke, T.~L.,
Wright, C.~M., Jorgensen, J.~K., van Dishoeck, E.~F., Wong, T., \apj,
in press (astro-ph/0306399)

\end{thebibliography}
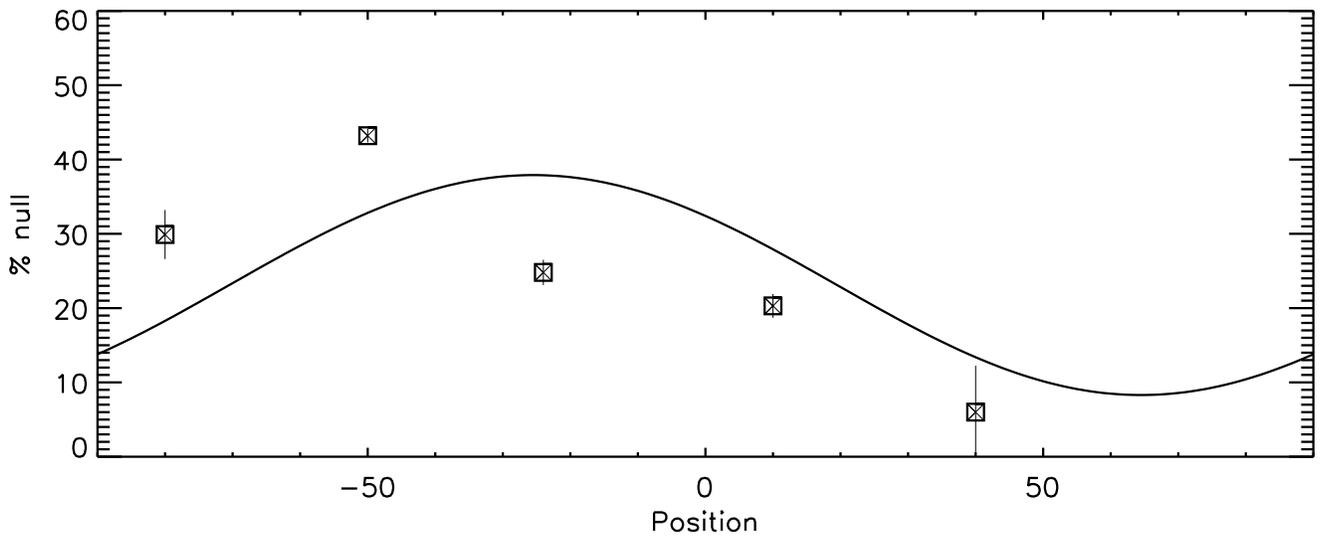
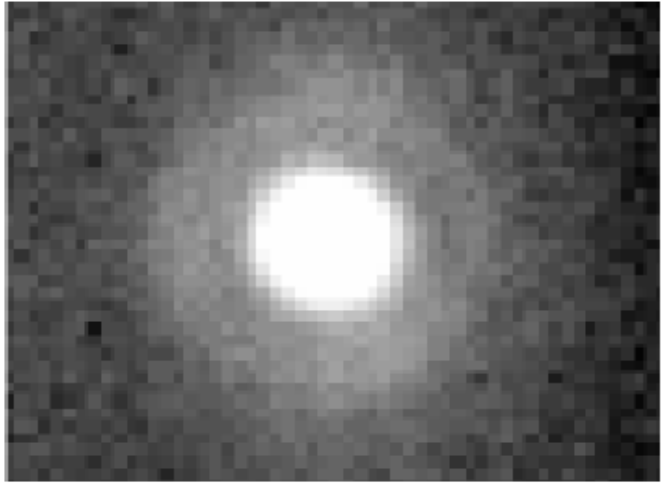
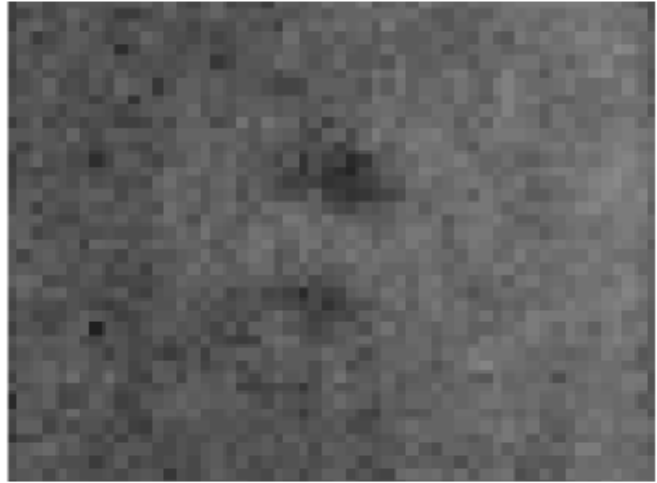
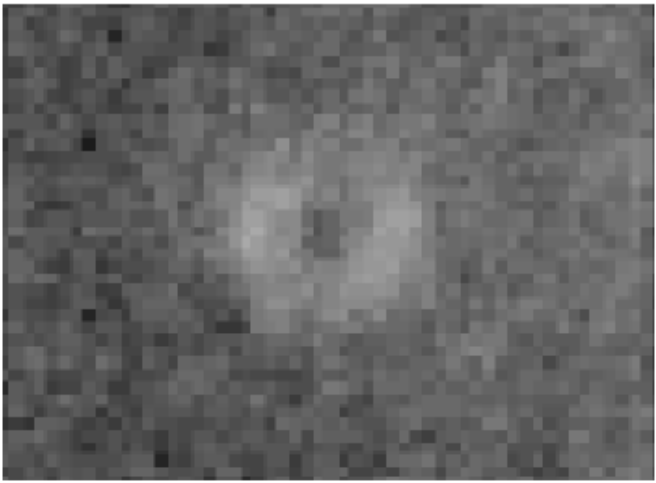
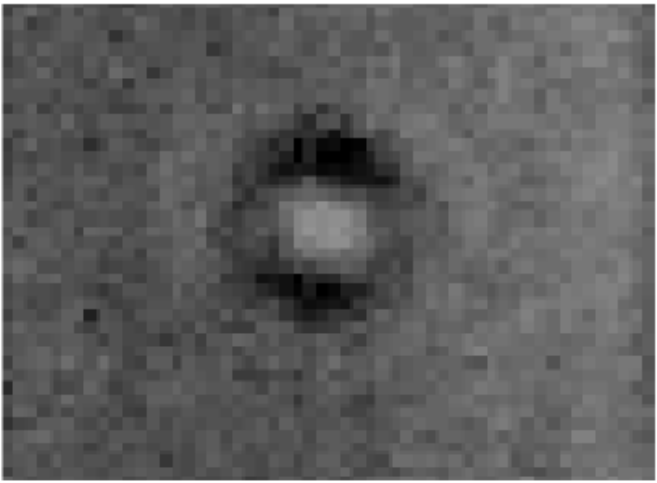
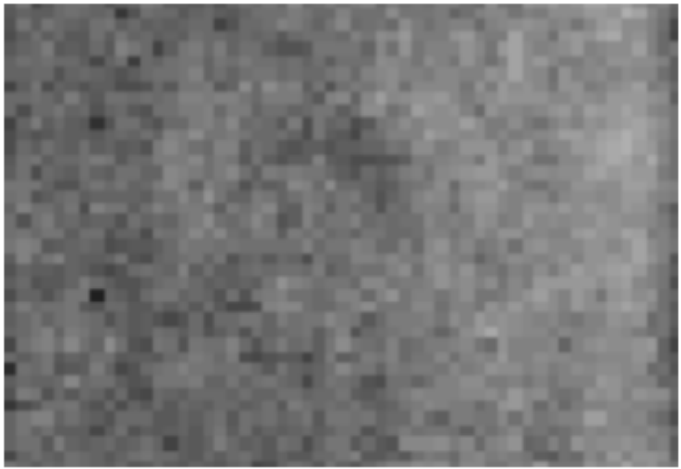
\end{document}